\documentclass{amsart} \usepackage{graphicx,pstricks,subfigure,hyperref}
\newtheorem{thm}{Theorem}[section] 
 
  \theoremstyle{definition} \newtheorem{defn}[thm]{Definition} \theoremstyle{remark}  \numberwithin{equation}{section}

\begin{document} 
\title{An Ultradiscrete QRT mapping from Tropical Elliptic Curves} 
\author{Chris Ormerod}
\begin{abstract} 
Recently the area of tropical geometry has introduced the concept of the tropical elliptic group law associated with a tropical elliptic curve. This gives rise to a notion of the tropical QRT mapping. We compute the explicit tropically birational expressions that define the tropical QRT mapping for some arbitrary set of parameters. We consider this a new integrable ultradiscrete system. This also induces a new notion of tropical elliptic functions, different to the ultradiscretized elliptic functions.
\end{abstract}
\maketitle
Recently the area of tropical geometry has received quite a bit of attention \cite{Intro}. Much of the work in this direction has been developed in parallel to algebraic geometry \cite{tropelliptic}. Many classical algebraic geometrical results transfer in some manner to the tropical world. In particular, there has recently been an algebraic geometric interpretation of the group law on an elliptic curve \cite{grouplaw}.

The QRT systems were introduced in \cite{QRT}. These mappings are birational mappings which can be described in terms of the group law on an elliptic curve \cite{ElliptQRT}. The algebraic geometric interpretation allows us to characterize such mappings \cite{Tsuda}. We use the algebraic geometric interpretation considered in \cite{ElliptQRT,Tsuda} to define an analogous QRT mapping defined in terms of the tropical algebraic geometric framework. We use certain properties to derive explicit tropical birational expressions for the tropical QRT mapping. We consider this system integrable in the sense that the system possesses an invariant defined by the fact that its evolution is restricted to a tropical elliptic curve. 

The ultradiscretization process has been used to connect integrable cellular automata to integrable difference equations through a limiting process \cite{Ultradiscrete}. This limiting process brings a subtraction free difference equation of multiplicative type to an equation over on the semiring $S$. These mappings have their own sense of integrability defined in terms of ultradiscrete Lax pairs \cite{Corm1} and ultradiscrete singularity confinement \cite{Joshi_2005}. Through this process, one derivation of an ultradiscrete QRT mapping has been given and is considered integrable \cite{Nobe}. These systems described by tropical elliptic curves should be related in some manner to the ultradiscretized systems.

In section 1 we introduce the concept of a tropical elliptic curves, lines and the tropical group law. In section 2 we derive an explicit expression for the tropical birational mapping we will call the tropical QRT mapping. This motivates our main result, which is the definition of the tropical 
elliptic function.

\section{Introduction}
We define the max-plus semiring to be the set $\mathbb{R}\cup \{-\infty\}$ with operations of addition and maximum. These operations are often referred to as tropical multiplication and tropical addition. We may consider $S$ a multiplicative group and thus we may refer to subtraction as the tropical division. The ultradiscretization process (found in \cite{Ultradiscrete}) is often used to derive systems over $S$. It is process that brings a rational expression ,$f$, in variables $a_1, \ldots,a_n$ to an expression $F$ in new ultradiscrete variables $A_1,\ldots, A_n$ related to the old variables by the relation $a_i = e^{A_i/\epsilon}$ by the limiting process
\[
F(A_1, \ldots, A_n) = \lim_{\epsilon \to 0} \epsilon \log(e^{f(a_1,\ldots, a_n)})
\]
for example we have a correspondence between operations and their tropical equivalents given by
\begin{eqnarray*}
a + b \to \max(A,B) \\
a b \to A+ B\\
a/b \to A-B
\end{eqnarray*}
and one may use this process to derive an ultradiscrete QRT system over $S$ \cite{Nobe}. Using geometric interpretation however, one is required to look at tropical geometric theory that runs parallel to the algebraic geometric theory of QRT mappings. 

Tropical geometry is indeed a new area \cite{Intro}. There is no clear consensus between certain key concepts such as the definition of a variety. The merger of ultradiscrete integrable systems with tropical geometry is also relatively new \cite{Ormerod}. We give a definition of a curve that meets our purposes.
\begin{defn}
A tropical curve associated with a function $f(X)$ is the set of points in $X$ in which $f$ is not linear. We denote the curve associated with the function $f$ 
\[
V( f) = \{ x \in X \textrm{ such that $f(x)$ is not linear } \}
\]
\end{defn}
It is worth noting that there is another description of algebraic varieties that can be stated in terms of more traditional algebraic geometry. If one takes a field such as the algebraic functions in one variable, $K$, coupled with a non-archimedean valuation $\nu : K \to \mathbb{Q}$, defined to be the index at $0$, one finds that a coherent definition of a tropical variety can be stated in terms of the closures of the mapping of algebraic varieties over $K$ via $\nu$ \cite{Intro}. The definition we have stated is closely related to the more algebraic one in that the two are equivalent in certain cases. Other lifts and non-archimedean valuations have been used in the field of integrable systems \cite{Ormerod}.

For the purposes of this study, we shall restrict out attention to $S^2$. Our first object of interest is a line. We define a line to be the variety associated with a degree 1 curve. Given a degree $1$ mapping over $S^2$, $f = a\otimes x \oplus b\otimes y \oplus c$, it is clear that any tropical linear curve has the shape described in figure 1. \begin{figure}[!ht]
\begin{pspicture}(8,4)
\psline(0,2)(4,2)
\psline(4,0)(4,2)
\psline(4,2)(6,4)
\psdot(4,1)
\psdot(1,2)
\psdot(5,3)
\end{pspicture}
\caption{3 collinear points}
\end{figure}
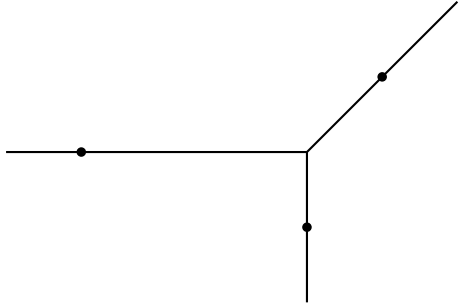
From this definition it is clear that any two generic points have a unique line through them, furthermore that every generic line intersects at one point. We define the vertex of the line to be the only point that is not linear. For a more concrete formalization of these concepts we must define the set of stable intersections, in which a tropical B\'ezout holds. For details about linear spaces we refer to \cite{linear}. A tropical elliptic curve however is 
\begin{defn}
A tropical elliptic curve is a smooth tropical curve of degree 3 and genus 1.
\end{defn}
which we state in the terminology of \cite{grouplaw} and \cite{linear}. The generic elliptic curve is then defined to be 
\begin{eqnarray}
\label{elli} V(f)&=&V(\max(a_0,x+a_1,y+a_2,2x+a_3,x+y+a_4,\\
&& 2y+a_5, 3x+a_6,2x+y+a_7,x+2y + a_8, 3y+a_9))\nonumber 
\end{eqnarray}
subject to the sufficient conditions of smoothness given by
\begin{eqnarray*}
a_0 > a_1 > a_3 > a_6\\
a_0 > a_2 > a_5 > a_9\\
a_1 > a_4 > a_8 \\
a_2 > a_4 > a_7 \\
a_3 > a_7 \\
a_5 > a_8
\end{eqnarray*}
with the exception that $a_0 = -\infty$ or $a_6 = -\infty$ or $a_9 = -\infty$ is permissable. Under these conditions, the curve is smooth and thus the group law holds. By following figure \ref{regions}, it is easy to construct a smooth elliptic curve of any specifications.
\begin{figure}[!ht]\label{regions}
\begin{pspicture}(10,9)
\psline(0,1)(1,1)
\psline(1,0)(1,1)
\psline(1,1)(2,2)
\psline(2,2)(2,5)
\psline(2,2)(6,2)
\psline(0,5)(2,5)
\psline(6,0)(6,2)
\psline(6,2)(8,4)
\psline(2,5)(3,6)
\psline(3,6)(8,6)
\psline(8,4)(8,6)
\psline(8,6)(10,8)
\psline(8,4)(9,4)
\psline(9,4)(10,5)
\psline(9,0)(9,4)
\psline(3,6)(3,8)
\psline(3,8)(4,9)
\psline(0,8)(3,8)
\rput(0.5,0.5){$a_0$}
\rput(3.5,0.5){$x + a_1$}
\rput(.5,3.){$y + a_2$}
\rput(7.5,1.5){$2x + a_3$}
\rput(5,4){$x + y + a_4$}
\rput(.5,6.5){$2y + a_5$}
\rput(10,2){$3x + a_6$}
\rput(9,5.5){$2x +y + a_7$}
\rput(6,7){$2y +x + a_8$}
\rput(1.5,9){$3y + a_9$} 
\end{pspicture}
\caption{The tropical elliptic curve under the conditions of \eqref{elli}. Each region is labelled according to which expression in the $\max$ is dominant. }
\end{figure}
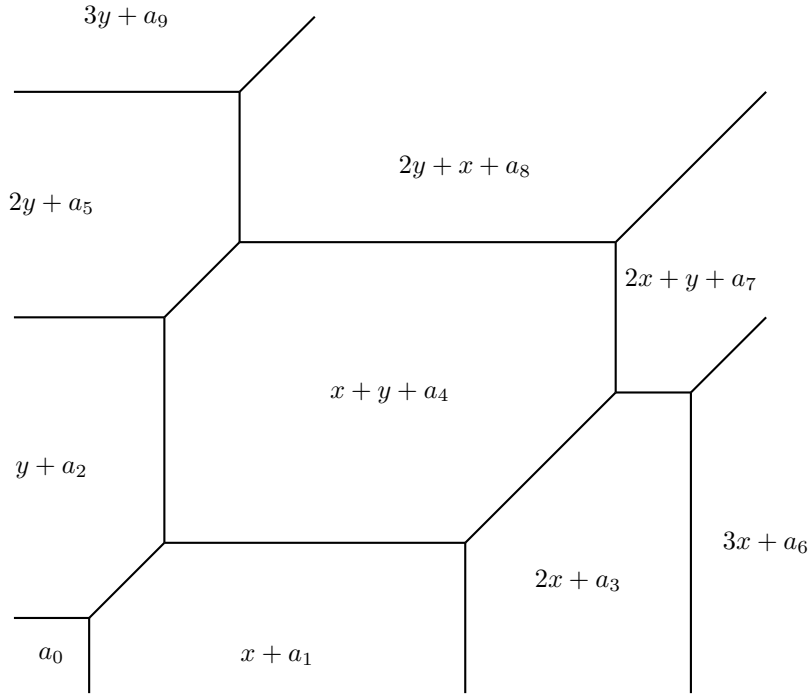
However, in order to define an elliptic curve $C = V(f)$, we need to consider the Jacobian. For the purposes of this article, it is sufficient to say the resultant construction shrinks the tentacles, which are the rays shooting out from the region where $x+y+a_4$ is dominant, to the boundary of the region where $x+y+a_4$ is dominant, which is also graphically represented by the thicker lines in figure 3. The details and formalization can be found in \cite{grouplaw} and references therein.

Graphically, it is clear that the line with a vertex inside the region will intersect in 3 points. By strengthening the condition that these be stable intersections, we note that any line whose vertex lies on the boundary of the elliptic curve also intersects the curve in three points. We simply define the addition by stating that the addition of these points is some constant point $\vartheta$ which is the tropical elliptic analog of the zero element. Figure 4 shows this relation.

Although, for generically chosen points, it is clear that there exists only one line through them, it is certainly not true that any two points define a single line. There are three cases, the case where the Euclidean line between the two points is parallel to the x-axis, y-axis and parallel to the line $y=x$. If the two points satisfy any one of the cases, then it is enough to say that the unique line through them is the one in which one of the points is the vertex of the line. 

In ordinary elliptic curves, there is a generic choice of point at $\infty$ to choose as the zero element, and on a Riemann sphere it is clear that this choice is more arbitrary. This choice gives a different group law, yet each gives us the necessary group. This is the same for the tropical analog of the group law in that we simply choose an arbitrary point on $C$ to be our $\vartheta$ element. 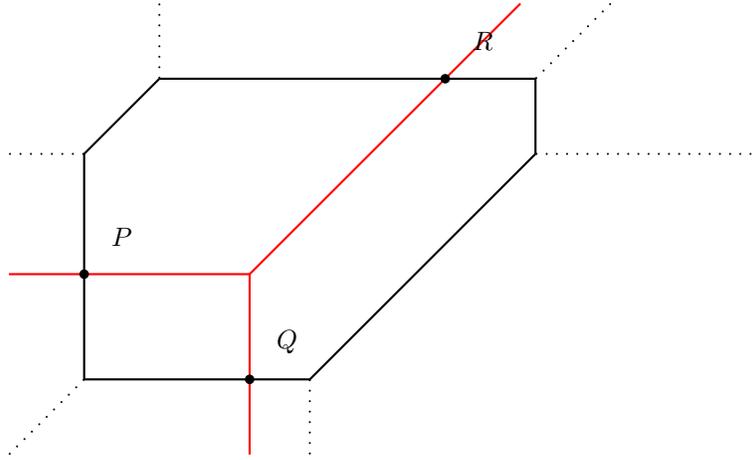
\begin{figure}[!ht]\label{grouplaw}
\begin{pspicture}(10,6)
\psline(1,1)(1,4)
\psline(1,1)(4,1)
\psline(1,4)(2,5)
\psline(4,1)(7,4)
\psline(2,5)(7,5)
\psline(7,4)(7,5)
\psline[linestyle=dotted](0,0)(1,1)
\psline[linestyle=dotted](7,5)(8,6)
\psline[linestyle=dotted](0,4)(1,4)
\psline[linestyle=dotted](4,0)(4,1)
\psline[linestyle=dotted](2,5)(2,6)
\psline[linestyle=dotted](7,4)(10,4)
\psline[linecolor=red](0,2.4)(3.2,2.4)
\psline[linecolor=red](3.2,0)(3.2,2.4)
\psline[linecolor=red](3.2,2.4)(6.8,6)
\psdot(1,2.4)
\psdot(3.2,1)
\psdot(5.8,5)
\rput(1.5,2.9){$P$}
\rput(3.7,1.5){$Q$}
\rput(6.3,5.5){$R$}
\end{pspicture}
\caption{A graphical representation of the addition law defined by $P+Q + R = \vartheta$}
\end{figure}
When considering the QRT mapping, the algebraic geometric interpretation of the map is one in which
\[
\overline{P} = P + T
\]
for some choice of $T$ on the elliptic curve. Within the literature, this point is usually chosen to be the point $(0,0)$. It is however unclear whether there is some generic point that we should choose in order to define our mapping. Figure 5 is a graphical representation of how we will define the QRT mapping for generically chosen $\vartheta$ and $T$, which will will assume are arbitrarily chosen.
\begin{figure}[!ht]\label{QRT}
\begin{pspicture}(10,6)
\psline(1,1)(1,4)
\psline(1,1)(4,1)
\psline(1,4)(2,5)
\psline(4,1)(7,4)
\psline(2,5)(7,5)
\psline(7,4)(7,5)
\psline[linestyle=dotted](0,0)(1,1)
\psline[linestyle=dotted](7,5)(8,6)
\psline[linestyle=dotted](0,4)(1,4)
\psline[linestyle=dotted](4,0)(4,1)
\psline[linestyle=dotted](2,5)(2,6)
\psline[linestyle=dotted](7,4)(10,4)
\rput(1.5,4){$\vartheta$}
\rput(3.6,.5){$P$}
\rput(.5,2.4){$T$}
\psline[linecolor=red](3.1,0)(3.1,2.1)
\psline[linecolor=red](0,2.1)(3.1,2.1)
\psline[linecolor=red](3.1,2.1)(7,6)
\rput(5.5,5.5){$-\overline{P}$}
\psline[linecolor=blue](0,3.5)(4.5,3.5)
\psline[linecolor=blue](4.45,3.5)(6.95,6)
\psline[linecolor=blue](4.5,0)(4.5,3.5)
\psdot(4.5,1.5)
\rput(5.5,1.2){$\overline{P} = P +T$}
\psdot(1,3.5)
\psdot(3.1,1)
\psdot(1,2.1)
\psdot(6,5)
\end{pspicture}
\caption{This is a tropical analog of the way in which the QRT mapping is defined geometrically.}
\end{figure}
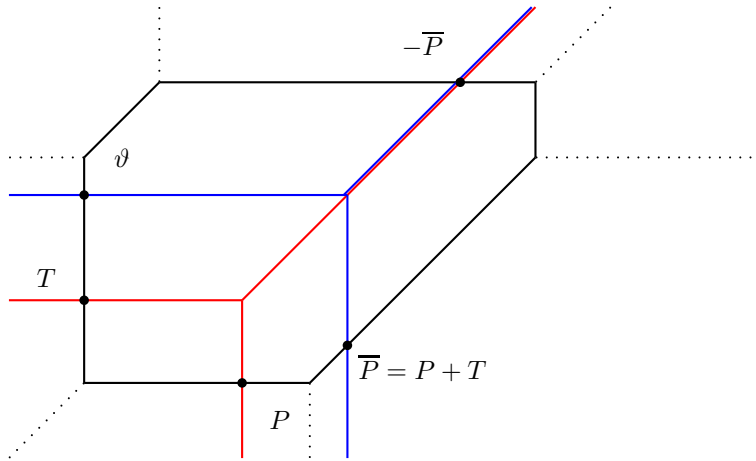

\section{Derivation}
Given two points, $P_0 = (x_0, y_0)$ and $P_1 = (x_1,y_1)$, if we wish to define find the formula for the unique line through these points. We also wish to do this in the manner described above in which the formula holds for points in general position, and also the case in which the points define lines parallel to the $x$-axis or $y$-axis or the line $y=x$ in the Euclidean sense. This line is simply given by the curve
\[
V( \max(x + \max(y_0,y_1), y + \max(x_0 , x_1), x_0 + y_1 , x_1 + y_0)
\]
It is clear that the vertex , $v = (v_x,v_y)$ of this line is at the point
\[
(v_x,v_y) = (\max(  x_0 + y_1 , x_1 + y_0) -  \max(y_0,y_1), \max(  x_0 + y_1 , x_1 + y_0) - \max(x_0 , x_1))
\] We see that any tropical line whose vertex lies inside the curve intersects any elliptic curve at three places, say $P$,$Q$ and $R$. These can be explicitly found given the vertex of the line. By dropping a line vertically down from the vertex, horizontally left from the vertex and parallel to $y=x$, we obtain 3 points given by
\begin{eqnarray*}
&(v_x, \max(a_1-a_4, v_x + a_3-a_4))\\
&(\max(a_2-a_4, v_y + a_5-a_4), v_y)\\
&(\min(a_4-a_7, v_x-v_y+a_4-a_8), \min(a_4-a_8, v_y-v_x +a_4-a_7))
\end{eqnarray*}
This gives us the basis for the tropical group law on the tropical elliptic curve. To find $P+Q = -R$ we simply add these three $x$ co-ordinates and subtract out the $x$ co-ordinates of $P$ and $Q$. Similarly, we may find an expression $y$ co-ordinate in terms of these points and the $y$ co-ordinates of $P$ and $Q$. By applying this twice, since $v_x$ and $v_y$ are expressions in the co-ordinates of $P$ and $Q$, and $\vartheta$ is some fixed point, we obtain a birational expression for $P + Q$.

We calculate explicitly the evolution of the QRT mapping when described in this manner. Let $T = (t_x,t_y)$, and the point $\vartheta = (z_x,z_y)$. We wish to calculate the iterate of $P = (p_x,p_y)$. We define the notation $P = P(n)$ and $\overline{P} = P(n+1)$. Using the formula for the generic 
line through the two points $T$ and $P$ we have the line
\[
V(\max(x + \max(t_y,p_y), y + \max(t_x,p_x), \max(p_x+t_y,p_y+t_x)))
\]
We note that this gives us the vertex of the line $v = (v_x,v_y)$ is given my the formula
\begin{subequations}\label{vu}
\begin{eqnarray}
v_x = \max(p_x+t_y,p_y+t_x)- \max(t_y,p_y)\\
v_y = \max(p_x+t_y,p_y+t_x)- \max(t_x,p_x)
\end{eqnarray}
\end{subequations}
We may calculate where these curves meet our elliptic curve giving the points
\begin{eqnarray*}
r_1 &=& (v_x , \max(a_1-a_4, v_x+a_3-a_4))\\
r_2 &=& (\max(a_2-a_4, v_y + a_5-a_4) , v_y)\\
r_3 &=& (\min(a_4-a_7, v_x-v_y+a_4-a_8), \min(a_4-a_8, v_y-v_x +a_4-a_7))\\
&=& (-\max(a_7-a_4,v_y-v_x+a_8-a_4), -\max(a_8-a_4, v_x-v_y+a_7-a_4))
\end{eqnarray*}
in which we may calculate the point $-\overline{P} = ((-\overline{p})_x,(-\overline{p})_y )$ from this by adding the three co-ordinates of the intersection of the line going through $v$ and subtracting the two co-ordinates given by $T$ and $P$. This gives us the formula for $-\overline{P}$ given by
\begin{subequations}\label{mp}
\begin{eqnarray}
(-\overline{p})_x &=& v_x + \max(a_2-a_4, v_y + a_5-a_4)- \\\nonumber && \max(a_7-a_4, v_y-v_x+a_8-a_4)- t_x - p_x\\
(-\overline{p})_y &=& v_y + \max(a_1-a_4, v_x + a_3-a_4)- \\ \nonumber && \max(a_8-a_4, v_x-v_y+a_7-a_4)- t_y - p_y
\end{eqnarray}
\end{subequations}
by using the formula for the line through generic points we note that the line through $-\overline{P}$ and $\vartheta$ is given by
\[
V(x + \max(z_y,(-\overline{p})_y) , y + \max(z_x, (-\overline{p})_x), \max(z_x + (-\overline{p})_y, z_y + (-\overline{p})_x))
\]
in which the center $u = (u_x, u_y)$ is given by
\begin{subequations}\label{uu} \begin{eqnarray}
\label{ux} u_x = \max(z_x + (-\overline{p})_y, z_y + (-\overline{p})_x))- \max(z_y,(-\overline{p})_y) \\
\label{uy} u_y = \max(z_x + (-\overline{p})_y, z_y + (-\overline{p})_x)) - \max(z_x, (-\overline{p})_x)
\end{eqnarray}
\end{subequations}
thus we may give a formula for the points given by the intersection of the tropical curve and the line. These are given by
\begin{eqnarray*}
r_1 &=& (u_x , \max(a_1-a_4, u_x+a_3-a_4))\\
r_2 &=& (\max(a_2-a_4, u_y + a_5-a_4) , u_y)\\
r_3 &=& (\min(a_4-a_7, u_x-u_y+a_4-a_8), \\ && \min(a_4-a_8, u_y-u_x +a_4-a_7))\\
&=& (-\max(a_7-a_4,u_y-u_x+a_8-a_4),\\&& -\max(a_8-a_4, u_x-u_y+a_7-a_4))
\end{eqnarray*}
now the formula for the point $\overline{P} = (\overline{p}_x, \overline{p}_y)$ is given by
\begin{subequations}\label{up}
\begin{eqnarray}
\label{pux}\overline{p}_x &=& u_x + \max(a_2-a_4, u_y + a_5-a_4) - \\
&&\max(a_7-a_4,u_y-u_x+a_8-a_4)- z_x - (-\overline{p}_x)\nonumber \\
\label{puy}\overline{p}_y &=& u_y + \max(a_1-a_4, u_x+a_3-a_4) - \\
&& \max(a_8-a_4, u_x-u_y+a_7-a_4) - z_y - (-\overline{p}_y)\nonumber 
\end{eqnarray}
\end{subequations}
The evolution can now be explicitly computed using (\ref{vu} - \ref{up}).
This now gives an explicit formula for the evolution defined by $\overline{P} = P + T$ for generically chosen curve, $\vartheta$ and $T$. This is then what we consider the analog of the QRT mapping for tropical systems. The resultant tropical equation however is birational of degree 6 in the original co-ordinates and is a particularly long expression, so will not be given here. However, we may give some particular examples. 

The evolution of this map to calculate the mapping for a particular example. The example we choose comes from the curve given by
\begin{equation}\label{example}
V(\max(x, y, x + y, 2x - 1, 2y - 1, 2x + y - 2, 2y + x - 2)) 
\end{equation}
which is given in figure 5. The evolution of a system defined on the curve \eqref{example} is computed and shown in figure 6.
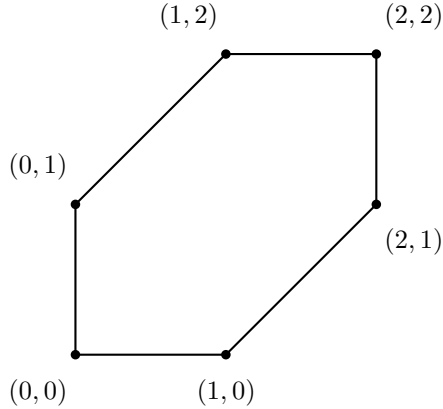
\begin{figure}[!ht]\label{eg}
\begin{pspicture}(6,6)
\psline(1,1)(1,3)
\psline(1,3)(3,5)
\psline(3,5)(5,5)
\psline(5,5)(5,3)
\psline(5,3)(3,1)
\psline(3,1)(1,1)
\psdot(1,1)
\psdot(1,3)
\psdot(3,5)
\psdot(5,5)
\psdot(5,3)
\psdot(3,1)
\rput(.5,.5){$(0,0)$}
\rput(.5,3.5){$(0,1)$}
\rput(2.5,5.5){$(1,2)$}
\rput(5.5,5.5){$(2,2)$}
\rput(5.5,2.5){$(2,1)$}
\rput(3,.5){$(1,0)$}
\end{pspicture}
\caption{The elliptic curve given by \eqref{example}}
\end{figure}

\begin{figure}[!ht]\label{exam}
\includegraphics[width = 12cm]{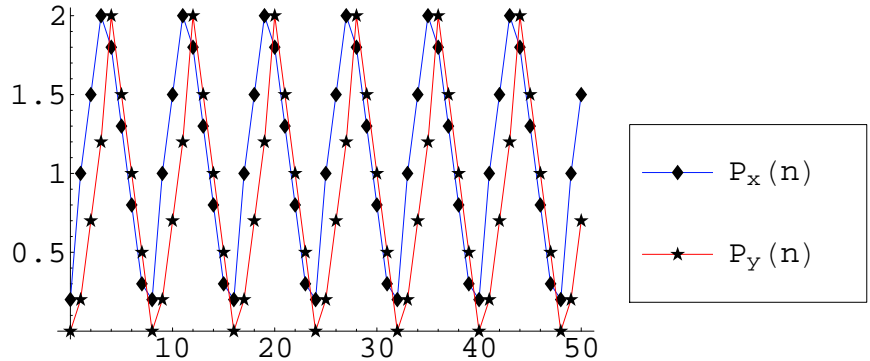}
\caption{The $QRT$ mapping associated with the curve given in figure \ref{eg} under the conditions $\vartheta = (0,0.5)$, $T = (0.5,0)$ and $P_0= (0.2,0)$.}
\end{figure}

We consider any particular solution a tropical elliptic function in the sense that the evolution is described in terms of tropical elliptic curves. Although we call this a tropical elliptic function, it would be interesting to know whether there is any correspondence between the functions defined in this manner and the ultradiscrete elliptic functions \cite{Nobe}. We conjecture that the ultradiscretized elliptic functions can by realized as particular solutions of the tropical elliptic functions.

\section{Conclusion}
This work warrants further investigation to QRT mappings as defined by this framework. It would be interesting to see whether there is some connection between mappings of this form and the ultradiscretized QRT mappings of the type given in \cite{Nobe}. Further question arise as to whether the tropical elliptic function in this sense define solutions of ultradiscrete equations such as the KP equation or ultradiscrete Painlev\'e equations.

\end{document}